\documentclass[aps,prd,nofootinbib,preprintnumbers,nobibnotes,reprint,twocolumn]{revtex4-1} 
\pdfoutput=1

\usepackage{amssymb}
\usepackage{amsmath}
\usepackage{graphicx,subfigure}
\usepackage{longtable}
\usepackage{verbatim}
\usepackage{amsfonts}
\usepackage{hyperref}
\usepackage{color}

\newcommand{\be}{\begin{equation}}
\newcommand{\ee}{\end{equation}}
\newcommand{\bea}{\begin{eqnarray}}
\newcommand{\eea}{\end{eqnarray}}
\newcommand{\beq}{\begin{equation}}
\newcommand{\eeq}{\end{equation}}
\def\beqa{\begin{eqnarray}}
  \def\eeqa{\end{eqnarray}}
\newcommand{\bv}{\left(\begin{array}{c}}
\newcommand{\ev}{\end{array}\right)}

\def\lsim{\mathrel{\rlap{\lower4pt\hbox{\hskip1pt$\sim$}}
    \raise1pt\hbox{$<$}}}	  
\def\gsim{\mathrel{\rlap{\lower4pt\hbox{\hskip1pt$\sim$}}
    \raise1pt\hbox{$>$}}}	  

\newcommand{\nn}{\nonumber}

\begin{document}

\title{ $\mathbf{ R(D^*)}$,  $\mathbf{|V_{cb}|}$, and the Heavy Quark Symmetry relations between form factors}
\author{Dante Bigi} 
\email{dante.bigi@to.infn.it}
\author{Paolo Gambino}
\email{gambino@to.infn.it}
\author{Stefan Schacht}
\email{schacht@to.infn.it}
\affiliation{
Dipartimento di Fisica, Universit\`a di Torino \& INFN, Sezione di Torino, I-10125 Torino, Italy}

\vspace*{1cm}

\begin{abstract}
Stringent relations between the $B^{(*)} \to D^{(*)} $ form factors exist in the heavy quark limit and the leading symmetry breaking corrections are known. We reconsider their uncertainty and role  in the analysis  of recent Belle data for $B\to D^{(*)}\ell\nu$ with model-independent parametrizations and in the related prediction of $R(D^{(*)})$. We find $|V_{cb}|=41.5(1.3) \ 10^{-3}$ and  $|V_{cb}|=40.6(^{+1.2}_{-1.3}) \ 10^{-3}$ using input from Light Cone Sum Rules, and $R(D^{*})=0.260(8)$. 
\end{abstract}

\maketitle

\section{Introduction}

Among the various flavour observables showing a significant deviation from their Standard Model (SM) predictions ({\it flavour anomalies}), 
those related to tree-level semileptonic $B$ decays have received remarkable attention, as they 
potentially represent  clean signals of New Physics. In addition to 
the  long-standing discrepancy between the determination of the CKM element $|V_{cb}|$ from inclusive and 
exclusive semileptonic $B$ decays, there is a $\sim 4\sigma$  anomaly \cite{Amhis:2016xyh} in the ratios of exclusive 
semileptonic decays to tau and to light leptons, 
\be
R(D^{(*)})= \frac{{\cal B}(B\to D^{(*)}\tau \nu)}{{\cal B}(B\to D^{(*)}\ell \nu)}.
\label{RD}
\ee
which could indicate a violation of lepton universality, and hence a clear departure from the SM. 
A full understanding of these relatively simple $B$ decays is a necessary condition to profit from the  potential of the Belle-II and LHCb experiments in the search for New Physics,
independently of these anomalies.

For what concerns the $B\to D$ channel, 
recent progress in the determination of the relevant form factors in lattice QCD \cite{Lattice:2015rga,Na:2015kha} and a new analysis of the $q^2$ spectrum in $B\to D\ell\nu$
by the Belle Collaboration \cite{Glattauer:2015teq} have resulted \cite{Bigi:2016mdz,Aoki:2016frl} in a
more precise value of $|V_{cb}|$ in reasonable agreement with the inclusive determination
\cite{Alberti:2014yda,Gambino:2016jkc} and in the precise prediction $R(D)=0.299(3)$.

The situation is not yet so favourable in the $B\to D^*$ channel, which has so far  
provided the most 
accurate exclusive determination of $|V_{cb}|$. First, unquenched lattice calculations 
 of the relevant form factor \cite{Bailey:2014tva,Harrison:2016gup}
are still limited to the zero-recoil point, where the $D^*$ is at rest in the $B$ rest frame. This implies that the experimentally measured shape, which vanishes at zero-recoil, must be {\it extrapolated}. 
Second, the experimental collaborations have generally performed this extrapolation using the Caprini-Lellouch-Neubert (CLN) parametrization~\cite{Caprini:1997mu}, and have published results in terms of the few parameters of this parametrization. Only recently, 
Belle has published a preliminary analysis \cite{Abdesselam:2017kjf} which, for the first time, includes deconvoluted kinematic and angular distributions,
without relying on a particular parametrization of the form factors. 

The new Belle results have allowed  for fits of the experimental spectra with different parameterizations, with surprising consequences on the resulting value of $|V_{cb}|$: while a CLN  fit leads to 
\be
 |V_{cb}|= (38.2\pm1.5) \ 10^{-3} ,\label{cln}\ee
in good agreement with the   Heavy Flavour Averaging Group (HFLAV) global average, $ |V_{cb}|= 39.05(75) \, 10^{-3}$ \cite{Amhis:2016xyh}, 
the fits performed with the Boyd-Grinstein-Lebed (BGL) parametrization \cite{Boyd:1997kz}  prefer a much higher value \cite{Bigi:2017njr,Grinstein:2017nlq,Jaiswal:2017rve},  
\be
 |V_{cb}|= (41.7\pm2.0) \ 10^{-3}, \label{bgl}\ee
 well compatible with the most recent inclusive result, $ |V_{cb}|= 42.00(63) \, 10^{-3}$ \cite{Gambino:2016jkc}.
 
As we have  emphasized in Ref.~\cite{Bigi:2017njr},  this strong dependence of $ |V_{cb}|$ on the parameterization  should be interpreted with great care because $i)$ it refers to a specific set of data and the large discrepancy between Eqs.\,(\ref{cln}) and (\ref{bgl}) may not carry on  to other  sets of data;
$ii)$  the physical information encoded in the 
CLN and BGL parametrizations are not equivalent.  Although they are grounded 
in the same foundations (analyticity, crossing symmetry, operator product expansion),
the CLN parametrization makes use of  Heavy Quark Effective Theory (HQET) relations
between the $B^{(*)}\to D^{(*)}$ form factors in various ways in order to reduce the number 
of independent parameters.  
Indeed, Heavy Quark Symmetry requires all of these form factors
to be proportional to the Isgur-Wise function, and the leading symmetry breaking corrections of $O(\alpha_s, \bar\Lambda/m_{c,b})$ are known \cite{Neubert:1993mb,Luke:1990eg,Neubert:1991xw}. However, the residual uncertainty is not negligible and 
should be taken into account in the analysis of experimental data.
There are also a few precise lattice QCD 
calculations which test and complement these relations and should be taken into account.

The main purpose of the present paper is to investigate to which extent the Heavy Quark Symmetry relations between the form factors affect the results of our previous analysis, once
their uncertainty is properly accounted for.
The methodology developed to this end will then be applied to the calculation of $R(D^*)$, where the only available information on the scalar form factor comes from the form factor relations.

In  determining the  uncertainty  of the HQET relations between form factors we will 
assume a rather conservative approach, as  we believe it is required in order to test the SM  in the current situation. 
It has already been emphasized \cite{Bernlochner:2017jka} that the CLN parameterization does not account for uncertainties in the values of the subleading Isgur-Wise functions (and their derivatives) at zero recoil obtained with QCD sum rules \cite{Neubert:1992wq,Neubert:1992pn,Ligeti:1993hw}, and also that  the impact of  higher order corrections cannot be neglected \cite{Bigi:2017njr}.  
While only a few of the relevant form factors have been computed in lattice QCD, they provide useful information to improve the ratios and estimate their uncertainty. 
Our first  task will therefore be to use the form factor ratios and their uncertainties in deriving 
{\it strong} unitarity bounds on the coefficients of the BGL parametrization.  We will then use these bounds directly in the fit to experimental data, without deriving 
a simplified parametrization like in Ref.\,\cite{Caprini:1997mu}. Finally, we will apply the results of our fits to the calculation of $R(D^*)$.
  
Our paper is organized as follows:  in the next Section we discuss the uncertainties 
due to higher order effects in the HQET relations between form factors. In Sec.~3 
 we compute  the {\it strong} unitarity bounds on the form factors that enter the 
 $B\to D^* \ell\nu$ decay rate taking into account their uncertainties. In Sec.~4 we discuss 
 our new fits to the preliminary Belle data which incorporate the strong unitarity conditions.
 In Sec.~5 we compute $R(D^*)$ and discuss its uncertainty. Section 6 contains a brief summary and our conclusions.

\section{Uncertainty of the relations between form factors}

As explained in Refs.\,\cite{Boyd:1997kz,Caprini:1997mu} the unitarity constraints on the parameters of the
$z$-expansion can be made stronger by adding other hadronic channels which couple to 
 $\bar c \Gamma b$ currents. While in general this would require 
non-perturbative information on each form factor,
 in the case of the $B^{(*)}\to D^{(*)}$ transitions
the form factors are all related by Heavy Quark Symmetry, which can be used to 
simplify the task. 
These transitions are described by a total of 20 helicity amplitudes, which 
provide an appropriate basis of form factors. In the following we will adopt 
the notation of   Ref.~\cite{Caprini:1997mu} -- see in particular Eqs.\,(A.3-A.6), in which all form factors 
reduce to the Isgur-Wise function $\xi(w) $ in the heavy quark limit.
In this notation $S_{1-3}$ couple with a scalar charm-bottom current, $P_{1-3}$ with a pseudoscalar current, $V_{1-7}$($A_{1-7}$)  with a vector (axial-vector) current.
For later convenience,  we also provide in Table \ref{tab:translation-dictionary} the relation with the notation of  Ref.\,\cite{Boyd:1997kz}.

The HQET ensures that  the form factor $F_i(w)$ (here  $w=v\cdot v'$, with $v$ and $v'$  the
four-velocities of the incoming and outgoing mesons) 
 can be expanded in inverse powers of the heavy quark masses and in $\alpha_s$, which 
 at the Next-to-Leading order (NLO) results in
\be
F_i(w)= \xi(w) \left[ 1+ c^i_{\alpha_s} \frac{\alpha_s}{\pi} + c^{i}_{b} \,\epsilon_b + c^i_{c}\,\epsilon_c+\dots\right], \label{nlo}
\ee
where $\epsilon_{b,c}= \overline{\Lambda}/2m_{b,c}$,  the strong coupling is typically evaluated 
at $\mu\sim \sqrt{m_c m_b}$ with $\alpha_s(\sqrt{m_c m_b})\approx 0.26$, and the dots represent higher order corrections. We recall that the Isgur-Wise function $\xi(w)$ is normalized to 1 at zero recoil, $\xi(1)=1$. Some of the ratios are known at $O(\beta_0 \alpha_s^2)$ but the extra  corrections are very small \cite{Grinstein:2001yg}.

\begin{table*}[t]
\begin{center}
\begin{tabular}{c|c|c}
\hline\hline
		   &
$B\rightarrow D$   & 
$B\rightarrow D^*$ \\\hline
$V, 1^-$  &\begin{minipage}{7cm} 
		$f_+ = \frac{1+r}{2 \sqrt{r}} V_1$
	    \end{minipage} 
	  &\begin{minipage}{7cm} 
		$g = \frac{1}{m_B\sqrt{r}}  V_4$
	  \end{minipage} \\
$A, 1^+$ &\begin{minipage}{7cm}  
		---
	\end{minipage}
	&\begin{minipage}{7cm} 
		\protect{\begin{align}
		f 		&= m_B \sqrt{r} (1 + w) A_1\,, \nn\\
		\mathcal{F}_1   &= m_B^2  (1 - r) \sqrt{r} (1 + w) A_5 \nn
		\end{align}}
	\end{minipage} \\
$S, 0^+$ &  $f_0^{\rm BGL} = m_B^2 \sqrt{r} ( 1 - r) (1 + w ) S_1$  
	 & ---  \\
$P, 0^-$ & --- 
	 & \begin{minipage}{7cm} 
	 $\mathcal{F}_2 = \frac{1 + r}{\sqrt{r}}  P_1 $ 	
	  \end{minipage}
	 \\
\hline\hline
		   &
$B^*\rightarrow D$ &  
$B^*\rightarrow D^*$ \\\hline
$V, 1^-$  &\begin{minipage}{7cm} 
		$\hat{g} = -\frac{1}{m_B^*\sqrt{ r}} V_5$
	  \end{minipage}
	  &\begin{minipage}{7cm} 
		\protect{\begin{align}
		V_{+0}  &= \frac{1}{m_{B^*}  \sqrt{r}} V_6\nn\\
		V_{++}  &= \frac{1}{m_{B^*}\sqrt{r}} V_7\nn\\
		V_{0+}  &= -\frac{1+r}{\sqrt{ 2 r}} V_2 \nn\\
		V_{00}  &= \frac{1+r}{2 \sqrt{r }} V_3 \nn
		\end{align}}
	  \end{minipage}\\\hline
$A, 1^+$ &\begin{minipage}{7cm} 
		\protect{\begin{align}
		\hat{f}             &= -m_{B^*} (1 + w)\sqrt{r}\,  A_2\nn\\
		\hat{\mathcal{F}}_1 &= m_{B^*}^2 (1+w)\left( 1-r \right) \sqrt{ r } A_6 \nn
		\end{align}}
	\end{minipage}
	&\begin{minipage}{7cm} 
		\protect{\begin{align}
		A_{++} &= - m_{B^*} \sqrt{r} (1+w) A_3 \nn \\
		A_{+0} &= m_{B^*} \sqrt{ r} (1 + w) A_4 \nn \\
		A_{0+} &= -m_{B^*}^2(1 -  r) \sqrt{ r} (1 + w)  A_7   \nn
		\end{align}}
	\end{minipage} \\\hline
$S, 0^+$  & ---
	 &\begin{minipage}{7cm} 
		\protect{\begin{align}
		S_{0+} &= m_{B^*}^2 \sqrt{r} (1 - r)(1 + w) S_2 \nn\\
		S_{00} &= -m_{B^*}^2  \sqrt{\frac{r}2}(1 - r) (1 + w) S_3 \nn 
		\end{align}}
	  \end{minipage}\\\hline
$P, 0^-$  & 
\begin{minipage}{7cm}
   \protect{\begin{align}
 		\hat{\mathcal{F}}_2 = -\frac{1+r}{\sqrt{r}} P_2\nn
		\end{align}}
		\end{minipage}
	  & \begin{minipage}{7cm}
   \protect{\begin{align}
 		P_{0+} = \frac{1+r}{\sqrt{r}} P_3\nn
		\end{align}}
		\end{minipage}
\\\hline\hline
\end{tabular}
\caption{  Relations between form factors in the BGL (left) and CLN (right) notation.
$r$ is the ratio of meson masses $m_{D^{(*)}}/m_{B^{(*)}}$ appropriate for each channel.
Notice that $f_0^{\rm BGL}$ differs from the more common notation $f_0=f_0^{\rm BGL}/(m_B^2-m_{D}^2)$, used,~{\it e.g.},~in Ref.~\cite{Bigi:2016mdz}.
\label{tab:translation-dictionary}} 
\end{center}
\end{table*}
We will follow here the calculation of  Ref.\,\cite{Bernlochner:2017jka} which updates 
those employed in the CLN paper. In particular, we  will adopt the  values of quark masses and of the subleading  parameters   given there,
\begin{align}
\eta(1)         &= 0.62\pm 0.20\,,  \qquad 
\eta'(1)        = 0.0 \pm 0.2\,, \nn \\
\hat{\chi}_2(1) &= -0.06 \pm 0.02\, \qquad \hat{\chi}'_2(1) = 0 \pm 0.02\nn\\
\hat{\chi}'_3(1) &= 0.04 \pm 0.02\, .
\end{align}
As the Isgur-Wise function cancels out in the ratios of form factors, the latter can be 
computed more accurately in the heavy quark expansion.
The central values of the ratios of form factors $F_j$ to $V_1$ computed in this way and expanded in $w_1=w-1$, 
\begin{align}
\frac{F_j(w)}{V_1(w)} = &A_j \left[1 + B_j \,w_1 +C_j \,w_1^2 + D_j \,w_1^3 + \dots \right],
\label{eq:update}
\end{align}
 are given in Table~\ref{tab:CLNtable-update},
which updates Table A.1 of Ref.~\cite{Caprini:1997mu} and has very similar results.

There is no obvious way to estimate the size of the higher order corrections
to the NLO HQET expressions.  Parametrically they are $O(\alpha_s^2)$, $O(\alpha_s \epsilon_c)$, and
$O(\epsilon_c^2)$, where  roughly 
\be
\alpha_s^2\sim  \alpha_s \epsilon_c \sim \epsilon_c^2\sim 6\%,
\ee
but the choice of $m_c$ or of the scale $\alpha_s$ can easily change this estimate.
Most importantly, the coefficients in front of these parameters 
can enhance or suppress significantly their contribution. For instance, the perturbative expansion is actually an expansion in $\alpha_s/4\pi$ and the two-loop is generally enhanced by $\beta_0\sim 9$. In the following  we will mostly worry about power corrections.

It is useful to recall that several of the form factors do not receive 
NLO power corrections at zero recoil because of Luke's theorem \cite{Luke:1990eg}. In particular, all of 
the scalar and axial-vector form factors, $S_{1-3}$ and $A_{1-7}$, do not 
receive $1/m$ corrections at zero recoil. There are also exact kinematic relations between the (pseudo)scalar and (axial)vector form factors
at maximal recoil $w=w_{max}$ (corresponding to $q^2=0$)
 \bea
 S_{1,2,3}(w_{max})&=&V_{1,2,3}(w_{max}),\nn\\
   P_{1,2,3}(w_{max})&=&A_{5,6,7}(w_{max}) \label{wmax}
 \eea
which  introduce a link between form factors protected by Luke's theorem and others
which are not protected. As an effect, the $1/m$ corrections tend to be smaller
in $V_{1,2,3}$ and $P_{1,2,3}$ as well, as a sort of {\it indirect} Luke's protection.
Finally, there are the following exact relations between form factors at zero-recoil ($w=1$):
\bea 
&& S_2(1)=S_3(1),\nn \\
&& A_1(1)=A_5(1), \nn\\
&& A_2(1)=A_6(1),  \label{zerorecoil}\\
&& A_3(1)=A_4(1)=A_7(1) .  \nn 
\eea
In some cases, such as $V_1$, the leading power corrections at zero-recoil are known to be
 suppressed  \cite{Uraltsev:2003ye}.
 \begin{table}[t]
\begin{center}
\begin{tabular}{c|r|r|r|r}
\hline\hline
 $F_j$  & $A_j$ & $B_j$ & $C_j$ & $D_j$   \\\hline
 $S_1$  & 1.0208 	& $-0.0436$ 	& $0.0201$	& $-0.0105$   \\
 $S_2$  & 1.0208 	& $-0.0749$ 	& $-0.0846$	& $0.0418$   \\
 $S_3$  & 1.0208 	& $0.0710$ 	& $-0.1903$	& $0.0947$   \\
 $P_1$  & 1.2089 	& $-0.2164$ 	& $0.0026$	& $-0.0007$   \\
 $P_2$  & 0.8938 	& $-0.0949$ 	& $0.0034$	& $-0.0009$   \\
 $P_3$  & 1.0544 	& $-0.2490$ 	& $0.0030$	& $-0.0008$   \\
 $V_1$  & 1 	& 0 	& 0	& 0   \\
 $V_2$  & 1.0894 	& $-0.2251$     & $0.0000$	& 0.0000  \\
 $V_3$  & 1.1777  	& $-0.2651$ 	& $0.0000$	& 0.0000   \\
 $V_4$  & 1.2351 	& $-0.1492$ 	& $-0.0012$	& 0.0003   \\
 $V_5$  & 1.0399 	& $-0.0440$ 	& $-0.0014$	& 0.0004   \\
 $V_6$  & 1.5808 	& $-0.1835$ 	& $-0.0009$	& 0.0003   \\
 $V_7$  & 1.3856  	& $-0.1821$ 	& $-0.0011$	& 0.0003   \\
 $A_1$  & 0.9656 	& $-0.0704$ 	& $-0.0580$	& 0.0276   \\
 $A_2$  & 0.9656 	& $-0.0280$ 	& $-0.0074$	& 0.0023   \\
 $A_3$  & 0.9656 	& $-0.0629$ 	& $-0.0969$	& 0.0470   \\
 $A_4$  & 0.9656 	& $-0.0009$ 	& $-0.1475$	&  0.0723  \\
 $A_5$  & 0.9656  	& $0.3488 $	& $-0.2944$	& 0.1456   \\
 $A_6$  & 0.9656 	& $-0.2548$ 	& $0.0978$ 	& $-0.0504$    \\
 $A_7$  & 0.9656 	& $-0.0528$ 	& $-0.0942$ 	& 0.0455    \\\hline\hline
\end{tabular}
\caption{Coefficients of the expansion in powers of $(w-1)$ of $F_j/V_1$, see Eq.\,(\ref{eq:update}).
 \label{tab:CLNtable-update}} 
\end{center}
\end{table}

The actual pattern of the NLO HQET corrections 
reflects these two qualitative suppressions: the form factors protected directly or  indirectly by Luke's theorem receive  small or moderate power corrections over the whole $w$ range
($1\le w\lsim 1.59(1.51)$ for $B\to D^{(*)}$),
 while the others ($V_{4,5,6,7}$) are affected by leading  power corrections as large as 50\%. 
The magnitude of the coefficients of $\epsilon_{b,c}$ reaches 2.1. The total NLO correction is almost 60\% in $V_6/V_1$, see Table \ref{tab:CLNtable-update}.
As Luke's theorem does not protect the form factors from $1/m^2$ corrections, it is therefore {\it natural}
to expect $1/m^2$ corrections of order 10-20\%, and one cannot exclude that  occasionally they can be even larger. 

The comparison with recent lattice QCD results is  instructive, even though it is 
limited  to the  few cases for which the form factors have been computed at zero or small 
recoil. Considering only unquenched lattice  results, we 
average those by the Fermilab/MILC and HPQCD collaborations \cite{Bailey:2014tva,Lattice:2015rga,Na:2015kha,Harrison:2016gup}, neglecting correlations between their results. We also mention that there is some tension between the preliminary value of $A_1(1)=0.857(41)$ by  HPQCD and the result of Fermilab/MILC, $A_1(1)=0.906(13)$. Incidentally we note that the first value agrees well with the heavy quark sum rule estimate
of Ref.\,\cite{Gambino:2012rd}.
The results at or near zero recoil are 
\begin{align}
S_1(w) &= 1.027(8) - 1.154(32)  (w-1)+\dots \nn\\
V_1(w) &= 1.053(8) - 1.236(33)  (w-1)+\dots , \label{lat0}\\
A_1(1) &= 0.902(12) \,,\nn
\end{align}
from which it follows that 
\begin{align}
& \frac{S_1(w)}{V_1(w)}\Big|_{\rm LQCD} =  0.975(6) +0.055(18) (w-1)+\dots, \nn\\
& \frac{A_1(1)}{V_1(1)}\Big|_{\rm LQCD} =  0.857(15), \label{lat}\\
& \frac{S_1(1)}{A_1(1)}\Big|_{\rm LQCD} =  1.137(21).\nn
 \end{align}
Notice that in the case of $S_1/V_1$ both numerator and denominator have been computed
at small recoil by the Fermilab/MILC and HPQCD collaborations, and we  therefore 
have also a lattice determination of the slope of the ratio. 

On the other hand, the  HQET calculation at NLO of Ref.\,\cite{Bernlochner:2017jka} gives  
\begin{align}
\left. \frac{S_1(w)}{V_1(w)} \right|_{\text{HQET}} &= 1.021(30) - 0.044(64) (w-1) +\dots\nn \\
\left. \frac{A_1(1)}{V_1(1)} \right|_{\text{HQET}} &= 0.966(28) \label{ratiohqe}\\
\frac{S_1(1)}{A_1(1)}\Big|_{\rm HQET} &=  1.055(2), \nn
\end{align}
where the errors represent only the parametric uncertainty on $m_b$, $\alpha_s$ and the QCD sum rules parameters.

Comparing the zero-recoil values of the ratios in Eqs.(\ref{lat}) to those in Eqs.\,(\ref{ratiohqe}) one 
observes deviations between  5\% and 13\%, which are obviously due to higher order  corrections unaccounted for in Eq.~(\ref{ratiohqe}). In all cases the deviation is larger than the NLO correction.
While it is quite possible that lattice 
uncertainties are somewhat underestimated, here we are not interested in a precision 
determination. What matters here is that the {\it size} of these deviations is 
consistent with our discussion above. 
The slope of the ratio $S_1/V_1$  computed on the lattice has a different sign from the one in (\ref{ratiohqe}) and their difference 
induces a 6\% shift at maximal $w$. However,  since $S_1/V_1=1$ at maximal recoil,   it is not surprising that higher order corrections are moderate in this case.

In conclusion,  higher order corrections to the form factor ratios computed in HQET at NLO
are generally sizeable and can  naturally  be of the order of 10-20\%.\footnote{
The CLN form factors $F_i$ we consider are 
helicity amplitudes which are linear combinations of the form factors in terms of which the matrix elements are decomposed. This sometimes leads to a correlation between numerator and denominator in the ratios  
of helicity amplitudes, which could affect our error estimates. There are only a few such cases among the ratios we employ in this paper: $S_1/V_1$, $P_1/A_1$, $A_5/A_1$, $P_1/A_5$.
The correlation is maximal at zero recoil where $A_1=A_5$ and one has
\begin{align}
\frac{S_1}{V_1} &= 1 + \frac{1-r}{1+r} \frac{h_-}{ h_+ - \frac{1-r}{1+r} h_- } = 1 + 0.48 \left(  0.04 \pm 0.06 \right)\,, \nn\\
\frac{P_1}{A_1} &= \frac{2}{1+r} - \frac{1-r}{1+r} \left( \frac{h_{A_2}}{h_{A_1}} + \frac{h_{A_3}}{h_{A_1}}\right) 
		= 1 - 0.45\left( -0.54 \pm 0.13 \right)\,.  \nn
\end{align}
Here we have also reported the NLO HQET result. In $S_1/V_1$ we have small
NLO corrections further suppressed by the prefactor $\approx 0.48$. This  
suggests that higher order corrections are somewhat suppressed, as in fact we found by comparing with LQCD above.  
In the second line the NLO corrections are sizeable despite the suppression factor, and also here one can naturally expect NNLO corrections between 10\% and 20\%.} 

\section{Strong unitarity bounds for $\mathbf{B\to D^*}$ form factors}

\begin{table*}[t]
\begin{center}
\begin{tabular}{c|c|c|c|c|c|c|c|c}
\hline\hline
$BD$      &    
$BD^*$    &    
$B^*D$   &    
$B^*D^*$ &  Type   & Mass (GeV)  &  Method & Decay const.(GeV) & Refs. \\\hline
\checkmark & \checkmark & \checkmark & \checkmark & $1^-$  & $6.329(3)$    & Lattice    & $0.422(13)$   &\cite{Olive:2016xmw, Dowdall:2012ab,Colquhoun:2015oha}\\
\checkmark & \checkmark & \checkmark & \checkmark & $1^-$  & $6.920(18)$   & Lattice    & $0.300(30)$   &\cite{Dowdall:2012ab,Ikhdair:2005xe}\\
\checkmark & \checkmark & \checkmark & \checkmark & $1^-$  & $7.020$       & Model              &               &\cite{Rai:2014fga}\\
$\times$   & $\times$   & $\times$   & \checkmark & $1^-$  & $7.280$       & Model              &               &\cite{Eichten:1994gt}\\\hline
$\times$   & \checkmark & \checkmark & \checkmark & $1^+$  & $6.739(13)$   & Lattice            &               &\cite{Dowdall:2012ab} \\
$\times$   & \checkmark & \checkmark & \checkmark & $1^+$  & $6.750$       & Model              &               &\cite{Godfrey:2004ya}\\
$\times$   & \checkmark & \checkmark & \checkmark & $1^+$  & $7.145$       & Model              &               &\cite{Godfrey:2004ya}\\
$\times$   & \checkmark & \checkmark & \checkmark & $1^+$  & $7.150$       & Model              &               &\cite{Godfrey:2004ya}\\\hline
\checkmark & $\times$   & $\times$   & \checkmark & $0^+$  & $6.704(13)$   & Lattice            &               &\cite{Olive:2016xmw, Dowdall:2012ab}\\
\checkmark & $\times$   & $\times$   & \checkmark & $0^+$  & $7.122$       & Model              &               &\cite{Godfrey:2004ya} \\\hline
$\times$   & \checkmark & \checkmark & \checkmark & $0^-$  & $6.275(1)$    & Experiment         & $0.427(6)$    &\cite{Olive:2016xmw,McNeile:2012qf} \\
$\times$   & \checkmark & \checkmark & \checkmark & $0^-$  & $6.842(6)$    & Experiment         &               &\cite{Olive:2016xmw}   \\   
$\times$   & \checkmark & $\times$   & \checkmark & $0^-$  & $7.250$         & Model            &               &\cite{Godfrey:2004ya} \\
\hline\hline
\end{tabular}
\caption{Relevant $B_c^{(*)}$ masses and decay constants, consistent with the subsets used in Refs.~\cite{Bigi:2016mdz,Bigi:2017njr}. 
We do not consider the fourth $1^-$ resonance as it is very close to threshold and its value is very uncertain. 
Predictions for the decay constant of the second $0^-$ resonance are also very uncertain and we do not include them here.
\label{tab:relevant-Bc-input}} 
\end{center}
\end{table*}

\begin{table}[t]
\begin{center}
\begin{tabular}{c|c}
\hline\hline
Input                   &     Value \\\hline
$m_{B^{*0}}$		&     5.325 GeV\\
$m_{B^0}$               &     5.280 GeV\\
$m_{D^{*+}}$            &     2.010 GeV\\
$m_{D^+}$		&     1.870 GeV	\\
$m_{\tau}$		&     1.77686 GeV \\
$\eta_{\mathrm{EW}}$    &  $1.0066$ \\
$\tilde\chi^T_{1^-}(0)$ &     $5.131\cdot 10^{-4}$ GeV$^{-2}$ \\
$\chi^L_{1^-}(0)$       &     $6.204\cdot 10^{-3}$\\
$\chi^T_{1^+}(0)$       &     $3.894\cdot 10^{-4}$ GeV$^{-2}$ \\
$\tilde\chi^L_{1^+}(0)$ 	&     $19.421\cdot 10^{-3}$ \\ 
\hline\hline
\end{tabular}
\caption{Additional numerical inputs. The calculation of the $\chi_i^{L,T}(0)$ follows Refs.~\cite{Grigo:2012ji,Bigi:2016mdz}.  $\chi^L_{1^-}$ and $\chi^L_{1^+}$ are needed for the scalar and pseudoscalar formfactors~\cite{Boyd:1997kz}; they are related to $\chi_{0^+}$ and $\chi_{0^-}$, respectively.
  \label{tab:further-input}} 
\end{center}
\end{table}

In the following we refer to the setup based on \cite{Boyd:1997kz}
which we have employed in \cite{Bigi:2017njr} to perform a fit to the recent Belle $B\to D^* \ell \nu$ differential distributions.
In this framework the generic  form factor $F_i$ (already in CLN notation) can be 
expressed as 
\be
F_i(w)= \frac{p_i(w) }{B_i(z)\phi_i(z)} \sum_{n=0}^{N} a_{n}^{(i)} z^n
\label{Fi}
\ee
where $z=(\sqrt{w+1}-\sqrt{2})/(\sqrt{w+1}+\sqrt{2})$ and the prefactors $p_i(w)$ 
are the ratios between helicity amplitudes in the CLN and BGL notations which 
can be read off Table~\ref{tab:translation-dictionary}. 
The series in $z$ in (\ref{Fi}) is truncated at power $N$ and we will set $N=2$ from the outset, which is sufficient at the present of level accuracy as  $0<z<0.056$ in the
physical region for semileptonic $B\to D^{*}$ decays to massless leptons. 

The Blaschke factors, $B_i(z)$,  take into account the subthreshold $B_c$ resonances with  
the same quantum numbers as the current involved in the definition of $F_i$. As the exact location of the threshold $(m_{B^{(*)}}+m_{D^{(*)}})^2$ depends on the particular $B^{(*)}\to D^{(*)}$ channel,   $B_i(z)$ may 
differ even between form factors with the same quantum numbers.
We will employ the resonances given in Table~\ref{tab:relevant-Bc-input}.
Finally, the {\it outer} functions $\phi_i(z)$ can be read from
Eq.(4.23) and Tables I and IV of Ref.\,\cite{Boyd:1997kz}
 and are given explicitly in a few cases in \cite{Bigi:2016mdz,Bigi:2017njr}. We will use $n_I = 2.6$ for the 
number of spectator quarks (three), decreased by a large and conservative SU(3) breaking 
factor. The other  inputs we use are given in Table \ref{tab:further-input} (all uncertainties are small and can be neglected), where the 
$\tilde\chi$ are the $\chi$ constants after taking into account the one-particle exchanges, see \cite{Caprini:1997mu,Bigi:2016mdz}.

Analyticity ensures that the coefficients of the $z$-expansion (\ref{Fi}) for the form factor $F_i$  satisfy the {\it weak} unitarity condition
\be
\sum_{n=0}^N (a_n^{(i)})^2 <1,\label{eq:weak}
\ee
but  there are a number of two body channels ($BD, BD^*,B^*D,B^*D^*, \Lambda_b \Lambda_c,\dots$) with the right quantum numbers, as well as higher multiplicity  channels, that give  positive contributions to the absorptive part 
to the two-point function and can  strengthen the unitarity bound on the coefficients of each  form factor. For instance,  the form factors
$A_{1,5}$  which appear in the $B\to D^*\ell\nu$ decays both contribute 
to the same unitarity sum with quantum numbers $1^+$,
\be
\sum_{n=0}^{N} \left(a_n^{A_1}\right)^2 + \left(a_n^{A_5}\right)^2 < 1.\label{weak}
\ee
However, the ({\it strong}) unitarity sums including all the $B^{(*)}\to D^{(*)}$ channels with quantum numbers $0^+$, $0^-$, $1^-$, $1^+$ are
\bea
\sum_{i=1}^3\sum_{n=0}^{N} \left(a_n^{S_i}\right)^2  < 1, \qquad
\sum_{i=1}^3\sum_{n=0}^{N} \left(a_n^{P_i}\right)^2  < 1,\nn\\
\sum_{i=1}^7\sum_{n=0}^{N} \left(a_n^{V_i}\right)^2  < 1, \qquad
\sum_{i=1}^7\sum_{n=0}^{N} \left(a_n^{A_i}\right)^2  < 1. \label{strong}
\eea
Now we can use the relations between the 20 form factors we have presented in the previous section to obtain constraints on the coefficients of {\it any} specific form factor $F_i$
\cite{Boyd:1997kz,Caprini:1997mu,Bigi:2016mdz}. It is sufficient to replace $F_j$ by $(F_j/F_i) \cdot F_i$ and expand the product in powers of $z$ to re-express each coefficient $a_n^{F_j}$ in terms of a linear combination of the $a_n^{F_i}$. 
In the case $N=2$ which is relevant here, each unitarity sum can then be reduced to a quadratic form in $a_0^{F_i}$,
$a_1^{F_i}$, $a_2^{F_i}$, and each unitarity condition in Eqs.\,(\ref{strong}) represents  an ellipsoid in the ($a_0^{F_i}$, $a_1^{F_i}$, $a_2^{F_i}$) space.

  \begin{figure*}[t]
 \begin{center}
  \includegraphics[width=7.5cm]{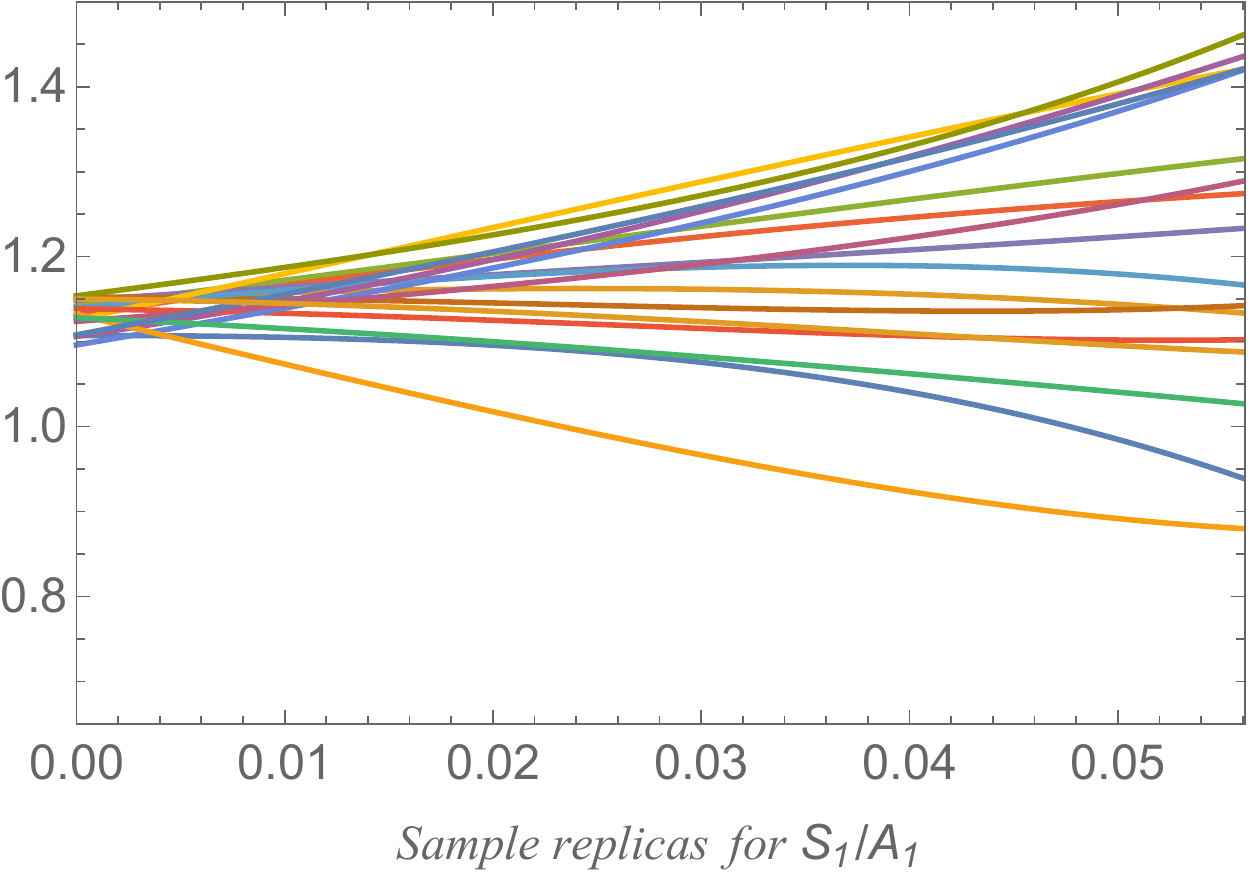}  \includegraphics[width=7.5cm]{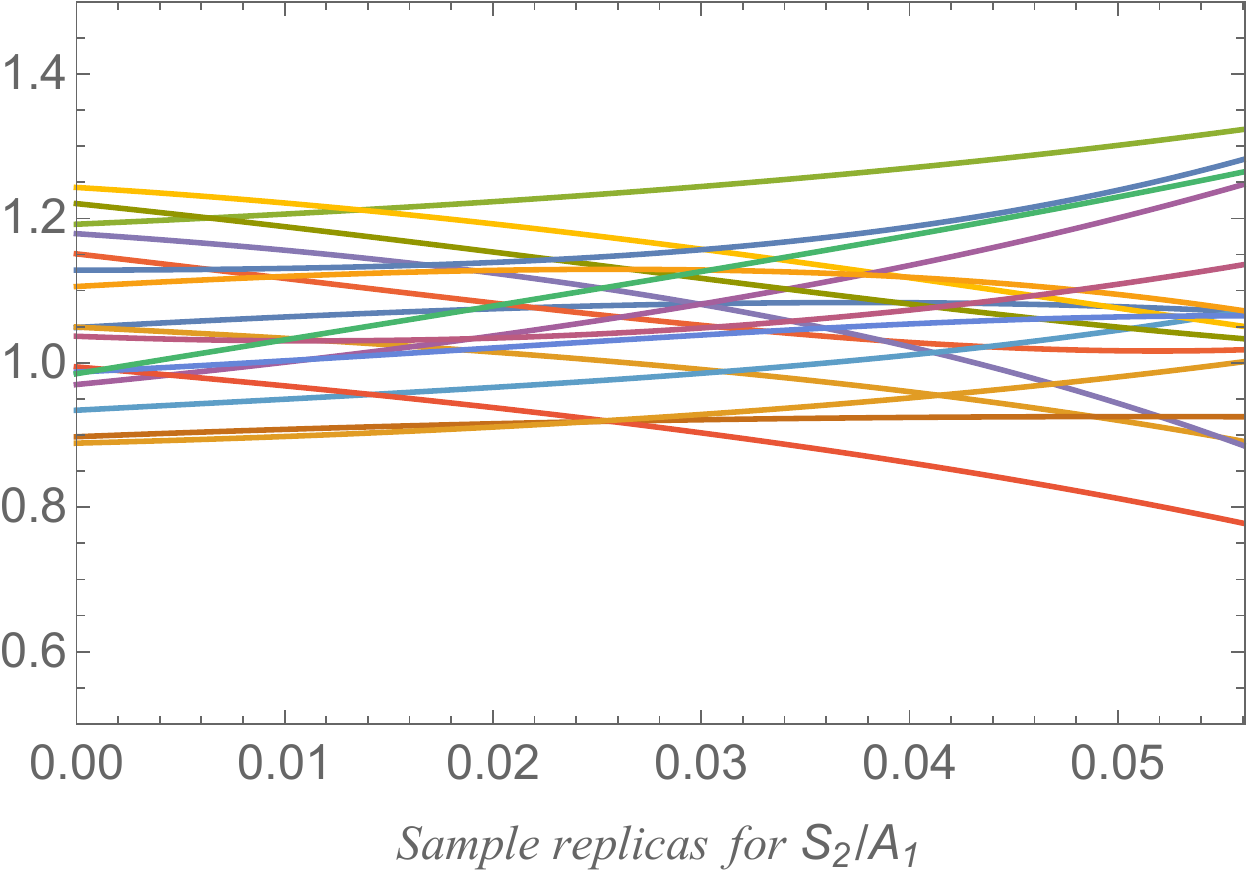}
  \caption{Sample replicas which satisfy all requirements for the form factor ratios 
  $S_{1,2}/A_1$ as a function of $z$. }
\label{fig:replicas}
\end{center}
\end{figure*} 

To take into account the uncertainties in the relations between form factors
we generate replicas of the set of ratios $F_j/F_i$ which satisfy the kinematic relations 
of Eqs.\,(\ref{wmax},\ref{zerorecoil}) and incorporate the lattice QCD results of 
Eqs.\,(\ref{lat}) within their uncertainties. Each replica must also have all of the 
ratios contained within a band 
around their central values computed at NLO in HQET as presented in the previous section, improved whenever possible with existing lattice data.
At zero-recoil the band has a width corresponding to the maximum between 25\% and 
$15\% +2\sigma_{HQET}$. At the endpoint, corresponding to $q^2=0$, the width 
is slightly larger and corresponds to  the maximum between 30\% and 
$20\% +2\sigma_{HQET}$. Here $\sigma_{HQET}$ is the total parametric 
relative uncertainty of the  NLO HQET calculation, obtained combining in quadrature the 
uncertainty from the QCD sum rule parameters, $m_b$, and $\alpha_s$.
   \begin{figure*}[t]
 \begin{center}
  \includegraphics[width=7.cm]{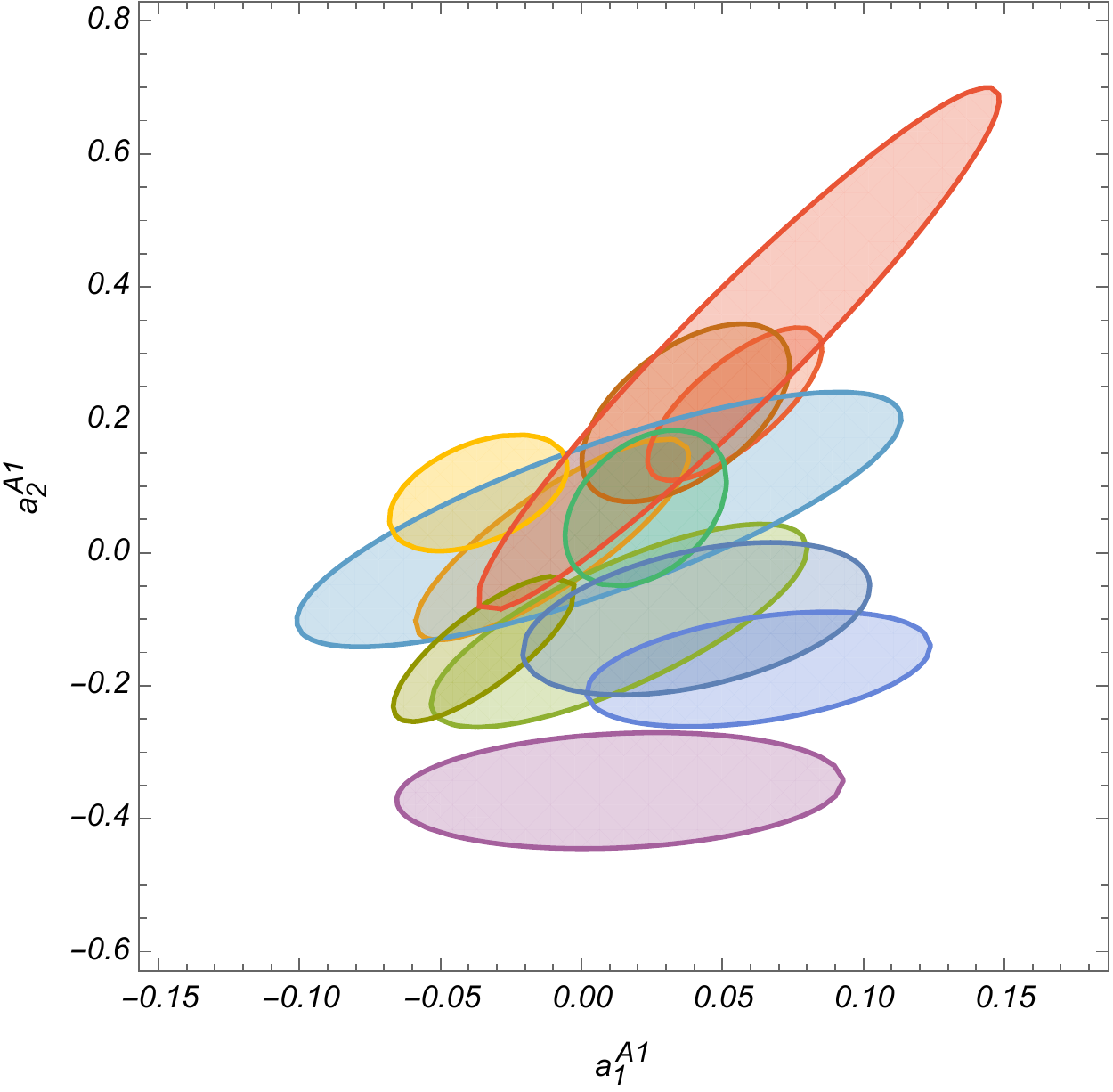}   \includegraphics[width=7.cm]{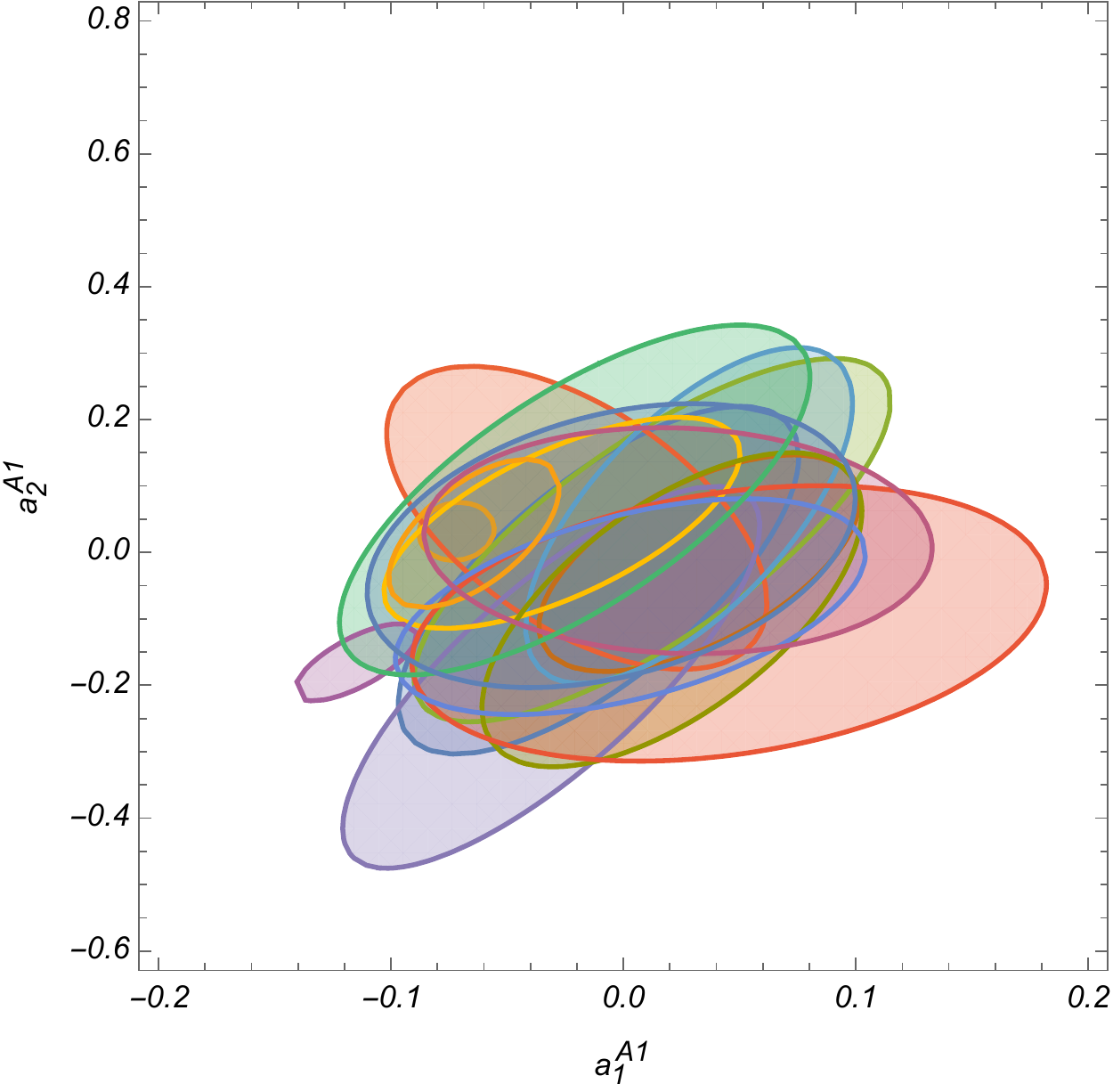}   \caption{Ellipses from sample replicas in the $(a_1^{A_1},a_2^{A_1})$ plane from the scalar and vector channels.  }
\label{fig:ellissi}
\end{center}
\end{figure*} 

Another condition that the  replicas $(F_j/F_i)_\alpha$ must comply with is that 
the coefficients $a_n^{(F_j)}$ of the form factor $F_j$, 
computed by expanding in powers of $z$  the expression 
\be
F_j(w) = (F_j/F_i)_\alpha  \, (F_i/V_1)_\alpha\,  V_1^{exp}(w)
\ee
satisfy weak unitarity, {\it i.e.} Eq.\,(\ref{eq:weak}).
 Here $V_1^{exp}(w)$ is the result of the fit to $B\to D\ell\nu$ experimental data  and lattice results 
 performed in \cite{Bigi:2016mdz}. The replicas $(F_j/F_i)_\alpha$
 are acceptable if unitarity is satisfied for values of coefficients of $V_1^{exp}(w)$
  within $3\sigma$s from their central values. 
Each replica therefore represents a viable model of the form factors.  
  An example of a few replicas of the ratios $S_{1,2}/A_1$  passing all tests
   is shown in Fig.~\ref{fig:replicas}.
   
\begin{figure*}[t]
 \begin{center}
  \includegraphics[width=5.9cm]{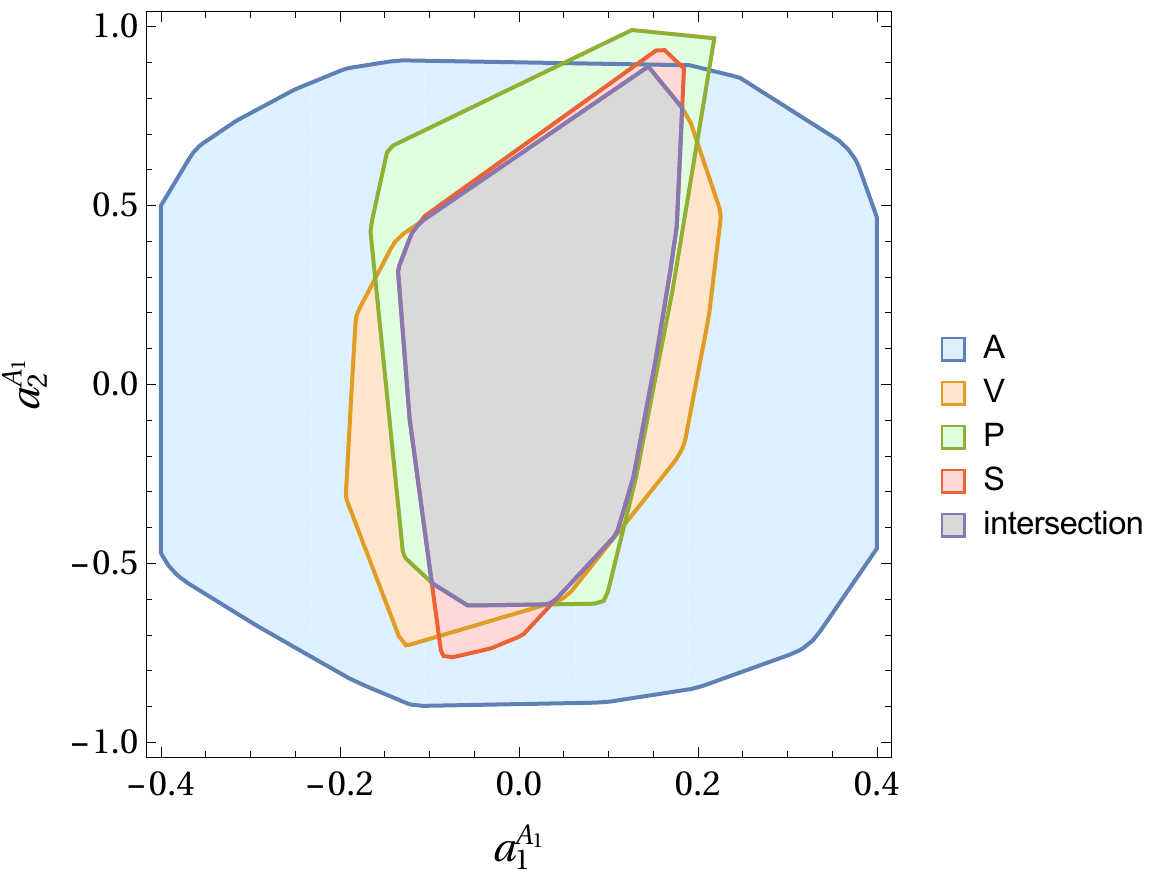}   \includegraphics[width=5.9cm]{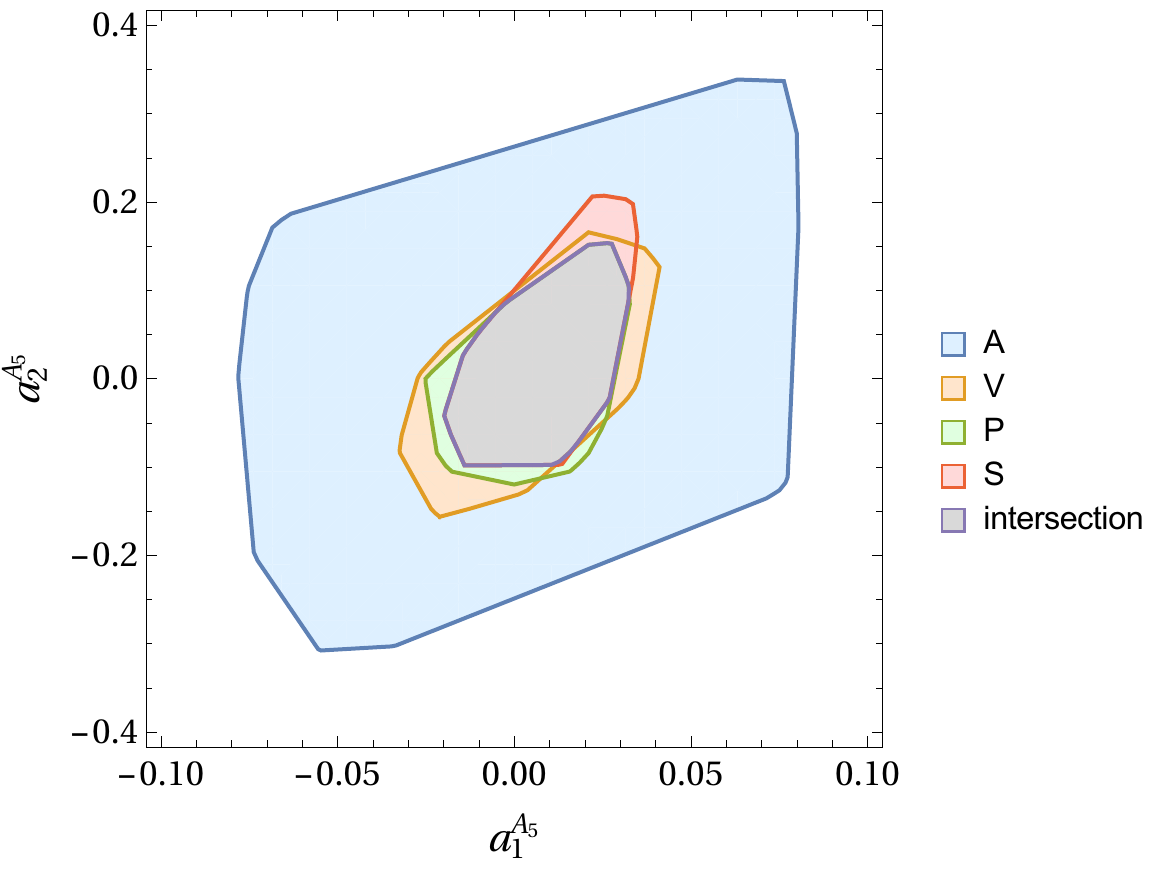}\\ 
   \includegraphics[width=5.9cm]{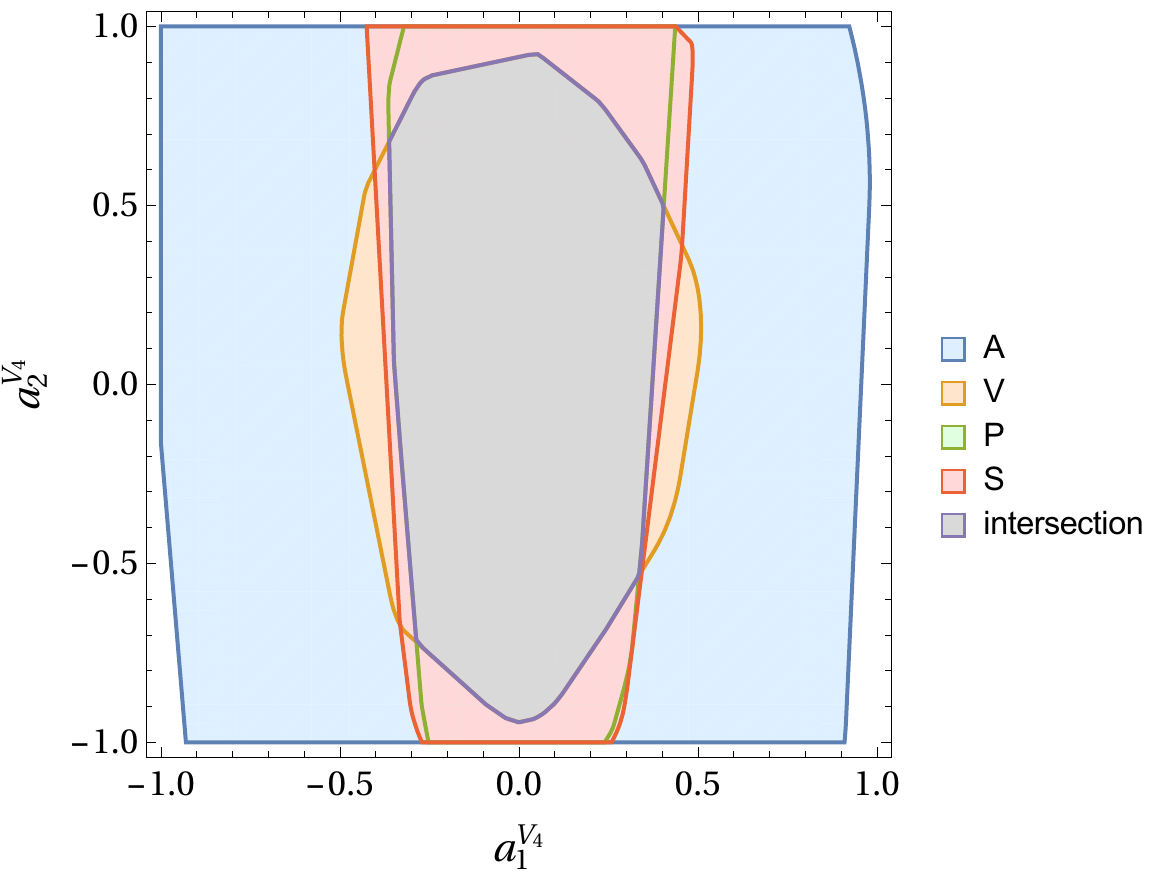}   \includegraphics[width=5.9cm]{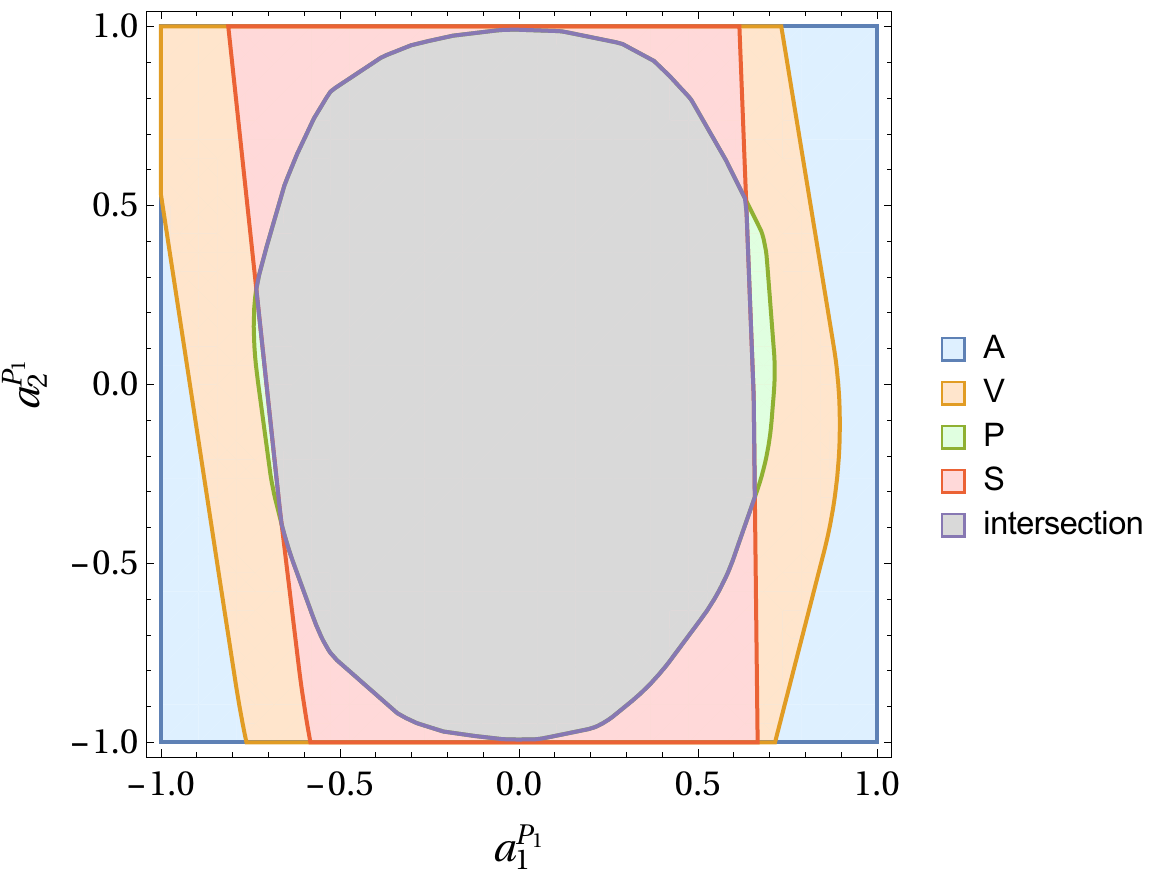} 
     \caption{Allowed regions in the 
   $(a_1^{A_1},a_2^{A_1})$, $(a_1^{A_5},a_2^{A_5})$,
 $(a_1^{V_4},a_2^{V_4})$ and $(a_1^{P_1},a_2^{P_1})$ planes from scalar (S), pseudoscalar (P), vector (V) and axial-vector (A) channels.}
\label{fig:envelopes}
\end{center}
\end{figure*}

We will be primarily interested in the four form factors which enter the 
$B\to D^* \ell \nu$ decays, namely $A_1,\ A_5$, $V_4$, and $P_1$.
The latter contributes only for massive final leptons.   
Each set of replicas of $F_i/A_1$ gives rise to four ellipsoids in the $(a_0^{A_1},a_1^{A_1},a_2^{A_1})$ space, corresponding to the four possible conditions in Eq.~(\ref{strong}). As $A_1(1)$ is known relatively well from lattice QCD calculations, 
we can fix $a_0^{A_1}$ and obtain 4 ellipses in the $(a_1^{A_1},a_2^{A_1})$ plane.
Samples of such ellipses from the S and V sectors are shown in Fig.~\ref{fig:ellissi}:
there is very little sign of correlation between $a_1^{A_1}$, $a_2^{A_1}$, but the regions identified in the two cases are similar. We have repeated the same procedure with a large 
number of replicas.  The envelope formed by all the ellipses represents 
the allowed region in the $(a_1^{A_1},a_2^{A_1})$ plane and is shown in Fig.~\ref{fig:envelopes}. It is quite remarkable that the allowed regions are very similar for the S, P, V channels, while the A channel is less constraining.
The intersection of the S, P, V and A channels is the allowed region we will consider in the following.

In the same way we have derived bounds in the $(a_1^{A_5},a_2^{A_5})$ plane. Indeed,
$A_5(1) $ is fixed by the same lattice calculations which fix $A_1(1)$.  The final results are also shown in Fig.~\ref{fig:envelopes}. The case of $V_4$ is slightly different because 
there is no lattice calculation that fixes $a_0^{V_4}$. In principle one should keep
the three-dimensional envelope of all ellipsoids. However, in line with the previous discussion, 
we can  assume that $V_4(1)$ is  within about 30\% from the values it takes when one uses  $A_1(1)$ or $V_1(1)$ together with HQET form factor ratios. This leads to  $0.0209<a_0^{V_4}<0.0440$. 
The bounds in the $(a_1^{V_4},a_2^{V_4})$
 plane depend little on the exact value of $a_0^{V_4}$ in that range, besides being anyway much weaker than those on the coefficients of $A_{1,5}$.
Therefore, also in this case we obtain a two-dimensional allowed region, shown in Fig.~\ref{fig:envelopes}.  The case of $P_1$ is very similar to that of $V_4$ and one similarly finds $0.041< a_0^{P_1}< 0.089$ and then a two-dimensional allowed region in the $(a_1^{P_1},a_2^{P_1})$.\footnote{The two-dimensional numerical regions are available from the authors upon request.}

Two comments are in order at this point.  
First, the weak or absent correlation between $a_1$ and 
$a_2$ that we observe in most cases does not imply the absence of a strong correlation 
between slope and curvature of the form factors when they are expressed in terms of the 
variable $w$. The latter  was observed long ago in Refs.~\cite{Boyd:1997kz,Caprini:1997mu} 
and  is a simple consequence of the change of variable from $z$ to $w$ and of the outer functions structure, combined with the weak unitarity bounds on $a_{1,2}$.
Indeed, if we proceed as in Ref.\,\cite{Caprini:1997mu},  we confirm their bounds on
slope and curvature of $V_1$.  The only
exception are  the constraints from the vector channel, which we find more constraining
than in  \cite{Caprini:1997mu}\footnote{We traced  the origin of the discrepancy to 
the  exponent of  $(\beta_j^2- (w+1)/2)$ in the denominator of the third row in their eq.(5)
(sum over $j=4-7$) which should be 4 instead of 5. The main results of \cite{Caprini:1997mu} are unaffected.}.

Second, we see no point in modifying the parametrization to include these stronger unitarity
bounds. The bounds we have found should  be used directly in fits to experimental and 
lattice data based on the BGL parametrization. In the  future, when new lattice information on the slopes of these form factors
will become available, the bounds can be simplified; they will  become one-dimensional bounds on $a_2^{F_i}$ only.

\section{Fits to $\mathbf{ B\to D^* \ell \nu}$ data}
We will now employ the results of the previous section in a fit to the available experimental data for $B\to D^*\ell \nu$, in order to illustrate the relevance of strong unitarity bounds 
in the present situation. To this end, we repeat here the analysis of Ref.\,\cite{Bigi:2017njr},
based on the preliminary Belle data of \cite{Abdesselam:2017kjf}, and refer to \cite{Bigi:2017njr} for all the details. 
The only additional piece of data we will include in the fit 
is the HFLAV average for 
the branching ratio of $\bar{B}^0\rightarrow D^{*+}l^-\bar{\nu}_l$ \cite{Amhis:2016xyh}
\begin{align}
\mathcal{B}(\bar{B}^0\rightarrow D^{*+}l^-\bar{\nu}_l) &= 0.0488\pm 0.0010\,, 
\label{BRhfag}
\end{align}
where we added the errors in quadrature. Combining it with 
 the total  lifetime $\tau_{B^0} = \left( 152.0\pm 0.4\right) \cdot 10^{-14} s$
 \cite{Olive:2016xmw}, we get a rather precise value for the total width of this decay.
 The above branching ratio can be compared with 
 \begin{align}
\mathcal{B}(\bar{B}^0\rightarrow D^{*+}l^-\bar{\nu}_l) &= 0.0495\pm 0.0025 \nn
\end{align}
reported in \cite{Abdesselam:2017kjf}. One would expect  the lower value in (\ref{BRhfag})
to drive the fit towards slightly lower values of $|V_{cb}|$ but we will see that 
the precision of the new input changes the fit in an unexpected way. 
We stress that the branching ratio is, to good approximation,  independent of the parametrization of the form factors used in the experimental analyses and it is therefore the only 
piece of data that we can use from older experimental results.
 We will also neglect all correlations of the total width with the binned angular and kinematic
distributions included in the fit. For what concerns the lattice determination of  the form factor at zero recoil, $A_1(1)$, we will use the average given in Eq.\,(\ref{lat0}), which differs slightly from the value employed in Ref.\,\cite{Bigi:2017njr}.

The results of the constrained fit are shown in Table \ref{tab:fit}, where we consider
fits in the BGL parametrization with weak and strong unitarity bounds, with and without the inclusion of the constraints  computed with Light Cone Sum Rules at $w=w_{max}$ in 
\cite{Faller:2008tr}: 
\bea
&A_1(w_{max})=0.65(18),
\label{LCSR}\\ &R_1(w_{max})=1.32(4),\quad
R_2(w_{max})=0.91(17).\nn
\eea
where
\bea
R_1(w)=\frac{V_4(w)}{A_1(w)} 
,\quad 
R_2(w)= \frac{w-r}{w-1}\left(1 -\frac{1-r}{w-r} \frac{A_5(w)}{A_1(w)}\right)
.\nn
\eea
\begin{table*}[t]
\begin{center} 
\begin{tabular}{c|c|c|c|c}
\hline
\hline
BGL Fit: 		 & Data + lattice 			& Data + lattice + LCSR & Data + lattice 			& Data + lattice + LCSR	\\\hline 
 unitarity & weak & weak & strong & strong \\ \hline
$\chi^2/\mathrm{dof}$    & $28.2/33$ 		& $32.0/36$	&$29.6/33$  &	 $33.1/36$		\\\hline
$\vert V_{cb}\vert$      & $0.0424\left(18\right)$ 	& $0.0413\left(14\right)$ 	& $0.0415\left(13\right)$  & $0.0406\left(^{+12}_{-13}\right)$	\\\hline

$a_0^{A_1}$ & $0.01218(16)$			  & $0.01218(16)$ 			& $0.01218(16)$ 	  & $0.01218(16)$			\\ 
$a_1^{A_1}$ & $-0.053\left(^{+56}_{-44}\right)$   & $-0.052\left(^{+25}_{-14}\right)$	& $-0.046(^{+34}_{-18})$  & $-0.029(^{+21}_{-13})$				\\
$a_2^{A_1}$ & $ 0.2\left(^{+8}_{-12}\right)$   & $0.99\left(^{+0}_{-46}\right)$	& $0.48(^{+2}_{-92})$ 	  & $0.5(^{+0}_{-3})$				\\\hline
$a_1^{A_5}$ & $-0.0101\left(^{+59}_{-55}\right)$  & $-0.0072\left(^{+52}_{-50}\right)$ 	& $-0.0063(^{+36}_{-11})$ & $-0.0051(^{+49}_{-13})$				\\
$a_2^{A_5}$ & $0.12\left(10\right)$    		  & $ 0.092\left(^{+92}_{-95}\right)$ 	& $0.062(^{+4}_{-64})$ 	  & $0.065\left(^{+9}_{-89}\right)$ 	\\\hline
$a_0^{V_4}$ & $0.011\left(^{+10}_{-8}\right)$     & $0.0286\left(^{+55}_{-36}\right)$ 	& $0.0209(^{+44}_{-0})$	  & $0.0299 (^{+53}_{-35})$			\\
$a_1^{V_4}$ & $0.7\left(^{+3}_{-4}\right)$     	  & $0.08\left(^{+8}_{-22}\right)$	& $0.33(^{+4}_{-17})$     & $0.04 (^{+7}_{-20})$			\\
$a_2^{V_4}$ & $0.7\left(^{+2}_{-17}\right)$	  & $-1.0\left(^{+20}_{-0}\right)$	& $0.6(^{+2}_{-13})$ 	  & $-0.9 (^{+18}_{-0})$			\\\hline\hline
\end{tabular}
\caption{Fit results using the BGL parametrization with $N=2$ 
 without and with the strong unitarity constraints. In the BGL fits $a_0^{A_5}$ is related to the value of~$a_0^{A_1}$, 
$a_0^{A_5}= 0.1675 \,a_0^{A_1}$.\label{tab:fit}}
\end{center} 
\end{table*}

We have also performed fits with the CLN parametrization (with free parameters $A_1(1), \rho^2, R_1(1), R_2(1)$) in the same way as in \cite{Bigi:2017njr}. We obtain 
$\vert V_{cb}\vert=0.0393(12) $  ($\chi^2/dof=35.4/37$) without the LCSR and $\vert V_{cb}\vert= 0.0392(12)$ ($\chi^2/dof=35.9/40$) with the LCSR. As expected, the difference between the values of $\vert V_{cb}\vert$ obtained with 
the BGL and CLN parametrization is reduced by the use of strong unitarity bounds, but it remains as large as  3.5-5\%, depending on whether LCSR results are included or not.

Comparing the fits in Table \ref{tab:fit} with those in Ref.\,\cite{Bigi:2017njr}
we note that the inclusion of the world average for the branching ratio has a significant impact 
on $\vert V_{cb}\vert$: the central value increases by 1.2 to 1.7\% and the error is
reduced by 10-20\%. Using the average of Eq.\,(\ref{lat0})
instead of the Fermilab/MILC result alone also leads to a minor increase of the $\vert V_{cb}\vert$ central value.

Comparing the fits in Table \ref{tab:fit}  with weak and strong unitarity bounds we observe that the strong constraints decrease $\vert V_{cb}\vert$ by 1.5-2.2\% 
and tighten its uncertainty quite a bit, especially in the less constrained fit without LCSR input.

It is also interesting to compare the effects of  the strong unitarity bounds we have derived
with the help of heavy quark symmetry relations with a naive  rescaling of the weak unitarity
conditions of Eq.~(\ref{weak}).
This gives an idea of  how strong the strong unitarity bounds really are and helps us
understanding their usefulness. The effects of the strong unitarity bounds is 
roughly similar to that of using 
\bea
&&\sum_{n=0}^2 (a_n^{V_4})^2 < \Lambda_V ,\nn\\
&&\sum_{n=0}^2 [(a_n^{A_1})^2+(a_n^{A_5})^2 ]< \Lambda_A \nn
\eea
with $\Lambda_{A,V}\lsim 0.2$, depending on the inputs. In effect, the strong unitarity bounds introduce little correlations among the $a_i^{F_j}$ coefficients: they mostly bound their individual size. This is unsurprising, as the unitarity sum rules cannot be saturated by one or two amplitudes only.

   \begin{figure*}[t]
 \begin{center}
  \includegraphics[width=8cm]{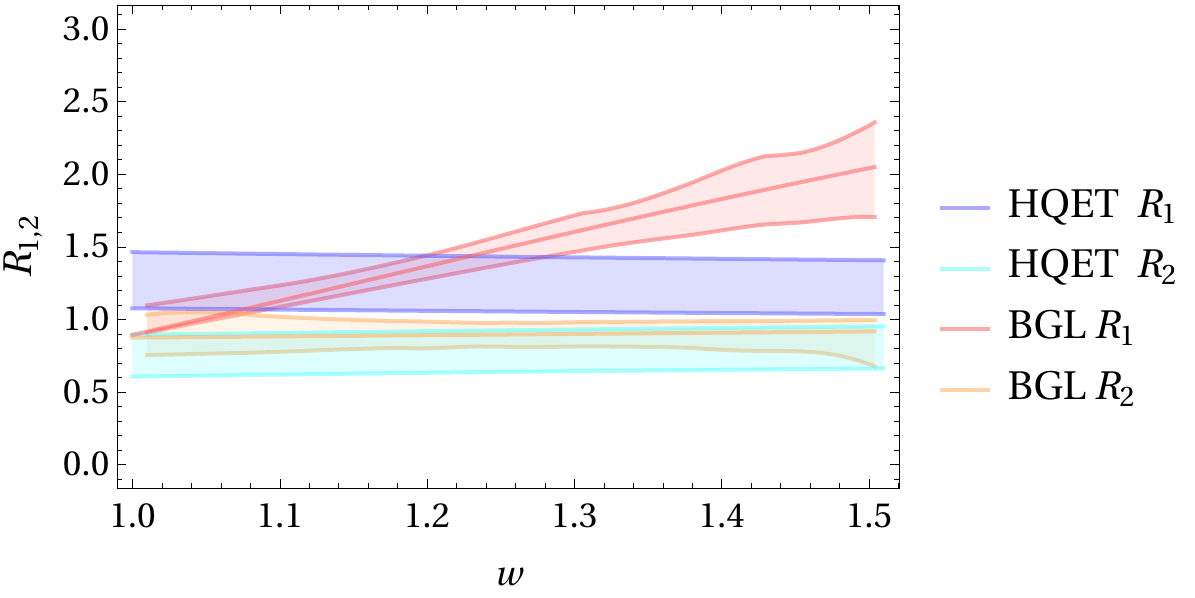}   \includegraphics[width=8.6cm]{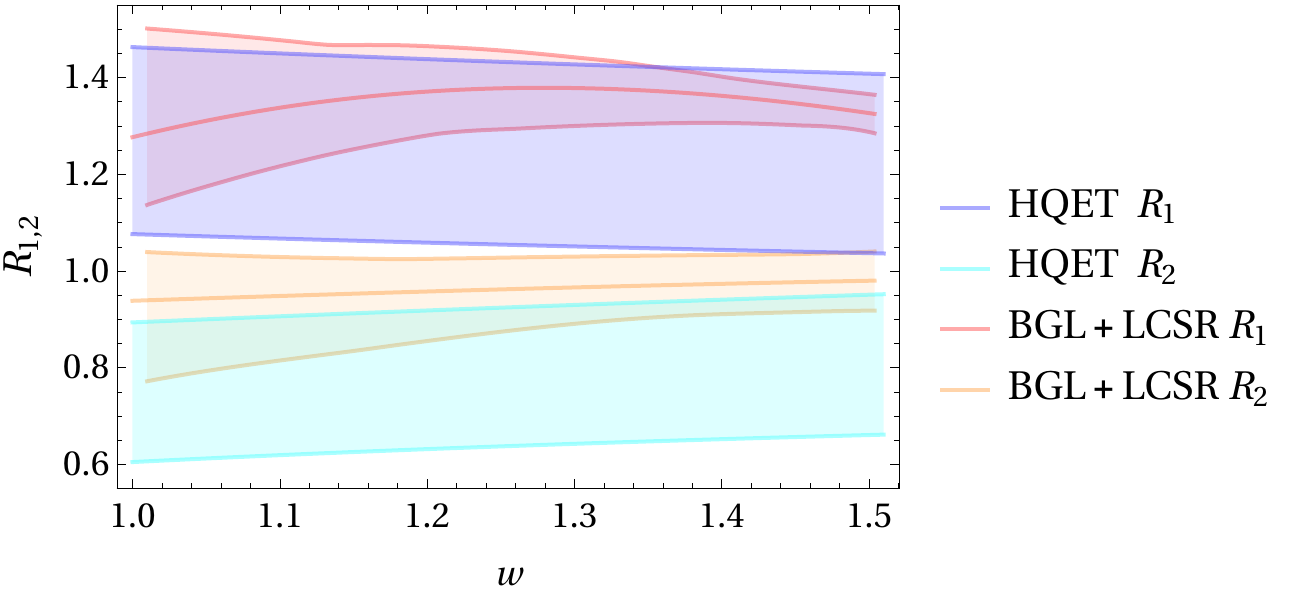}  
     \caption{Form factors ratios $R_{1,2}$ from the fits with strong unitarity bounds without (left) and with LCSR (right) input compared with their NLO HQET predictions, with parametric uncertainty combined in quadrature with a 15\% theoretical uncertainty.
          }
\label{fig:HQET}
\end{center}
\end{figure*} 
We now want to 
verify {\it  a posteriori} that the results of our fits are compatible with heavy quark symmetry within  reasonable uncertainties. 
Indeed, the form factor ratios $R_{1,2}(w)$ defined above after Eq.~(\ref{LCSR}) can be determined from the results of our fits.
A deviation from the NLO HQET predictions significantly larger than $\sim 20\%$ 
would signal an unexpected and unnatural breakdown of the heavy mass expansion.
This point has been emphasised in Refs.~\cite{Bigi:2017njr} and \cite{Bernlochner:2017xyx}.
The two plots in Fig.\,\ref{fig:HQET} show that the fits without/with LCSR lead to $R_2$ in good agreement with HQET (with input from QCD sum rules)
and the same holds for $R_1$ when LCSR are included.
On the other hand, without LCSR $R_1$ is well compatible with HQET
only at small or moderate recoil: at large $w$ there is a clear tension with both 
HQET and LCSR predictions. Lattice calculations will compute $A_1$ and $R_{1,2}$ 
at small recoil in the near future\footnote{Preliminary and incomplete results have been presented recently \cite{lattice2017}. They seem to exclude large deviations from HQET at small recoil.} and are likely to settle the whole $|V_{cb}|$ determination. In the meantime,
the fit without LCSR appears somewhat disfavoured.

Finally, we comment on the differences of our fits with those performed in \cite{Bernlochner:2017jka}. The main differences are that the authors of Ref.\,\cite{Bernlochner:2017jka}
employ the CLN parametrization for the reference form factor $A_1$ and 
assume the NLO HQET calculation for the form factor ratios $R_{1,2}$ without accounting for a theoretical uncertainty due to unknown higher order corrections. They also 
perform a combined fit to $B\to D$ and $B\to D^*$ Belle data.  
 In our fits we do not employ directly the HQET relations because we believe their present 
uncertainty does not make them useful and therefore a combined fit to  $B\to D$ and $B\to D^*$ data would give the same results of the two separate fits presented here and in  
\cite{Bigi:2016mdz}. Indeed, the coupling between the two sets of data through the unitarity bounds  would be extremely small.

\section{Calculation of $\mathbf{R(D^*)}$}
In the case of a massive ($\tau$) lepton the differential width for $B\rightarrow D^* \tau \nu_{\tau} $  can be written as the sum of two terms
\begin{align}
\frac{d\Gamma_{\tau}}{dw} &= \frac{d\Gamma_{\tau,1}}{dw} + \frac{d\Gamma_{\tau,2}}{dw} \label{eq:dGammadw-tau} \,.\nn
\end{align}
where
\begin{align}
\frac{d\Gamma_{\tau,1}}{dw} &= \left(1 -\frac{m_{\tau}^2}{q^2} \right)^2 
	\left(1 + \frac{m_{\tau}^2}{2 q^2}\right) \frac{d\Gamma}{dw} , \nn\\
\frac{d\Gamma_{\tau,2}}{dw} &=k \frac{ 
		 m_{\tau}^2  (m_{\tau}^2\! -\! q^2 )^2  r^3(1\!+\!r)^2 (w^2\!-\!1)^{\frac{3}{2}} P_1(w)^2
			}{
		 (q^2)^3
	} \,.\nn
\end{align}
Here $d\Gamma/dw$ represents the differential width for the  decay to massless 
leptons, see e.g.\ \cite{Bigi:2017njr}, and depends on the form factors $A_{1,5}$ and $V_4$.
 In the second term $k=\eta^2_{\mathrm{EW}} \vert V_{cb}\vert^2 G_F^2 m_B^5/32 \pi^3$, $r=m_{D^*}/m_B$, with  $\eta_{\rm EW}\simeq 1.0066$  the leading QED correction. 
The second term depends on a new form factor, $P_1(w)$, whose
 $z$-expansion  is
\be
P_1(w) = \frac{\sqrt{r}}{(1+r)B_{0^-}(z)\phi_{P_1}(z)}\sum_{n=0}^\infty a_n^{P_1} z^n \,. 
\label{P1exp}
\ee
where the outer function is given by
\begin{align}
\phi_{P_1} &=  \sqrt{\frac{n_I}{\pi \tilde{\chi}^L_{1^+}(0)}} \frac{8\sqrt{2}\, r^2 (1 + z)^2}{ 
	\sqrt{1-z} \left((1+r)(1-z) + 2 \sqrt{r} (1+z) \right)^4
	}  \nn 
\end{align}
The Blaschke factor $B_{0^-}(z)$  takes into account the first three $0^-$ resonances, see Table~\ref{tab:relevant-Bc-input}.  

The ratio $R(D^*)$, defined in Eq.\,(1),  can be split into two parts
\begin{align}
R(D^*)   &= R_{\tau,1}(D^*) + R_{\tau,2}(D^*)\,,\nn \\
R_{\tau,1}(D^*) &= \frac{
 	\int_1^{w_{\tau,\mathrm{max}}} dw\, d\Gamma_{\tau,1}/dw  
	}{
	\int_1^{w_{\mathrm{max}}} dw\, d\Gamma/dw  
}  \,,\\
R_{\tau,2}(D^*) &= \frac{
 	\int_1^{w_{\tau,\mathrm{max}}} dw\, d\Gamma_{\tau,2}/dw  
	}{
	\int_1^{w_{\mathrm{max}}} dw\, d\Gamma/dw  
}\,,
\end{align}
where 
\be
w_{\tau,\mathrm{max}}= (m_B^2 + m_{D^*}^2 - m_\tau^2)/(2 m_B m_{D^*})\approx 1.355.\label{wtaumax}
\ee

Unfortunately, the experimental $q^2$-spectrum of $B\rightarrow D^* \tau \nu_{\tau} $ cannot be reliably used to constrain the form of $P_1$ and there is no lattice calculation of this form factor.
We  therefore consider three options:
\begin{itemize}
\item to use $P_1= (P_1/A_1) A_1$ where $A_1(w)$ is taken from the fit and the ratio from HQET;
\item to use $P_1= (P_1/V_1) V_1$ where $V_1(w)$ is taken from the fit of \cite{Bigi:2016mdz} and the ratio from  HQET;
\item to use the HQET expression for $P_1(1)$ and the constraint $P_1(w_{max})=A_5(w_{max})$ together with unitarity. 
\end{itemize}
Having three alternative derivations will give us an additional 
handle to estimate  the overall uncertainty.
For a reference, we recall that the experimental world average for $R(D^*)$  is \cite{Amhis:2016xyh} (see update online)
\begin{align}
R(D^*)^{\mathrm{exp}} & =0.304 \pm 0.013\pm 0.007 \,.
\end{align}

\subsubsection{The standard way: normalizing $P_1$ to $A_1$}
The first option corresponds to the usual way of computing $R(D^*)$, see e.g. 
\cite{Fajfer:2012vx,Bernlochner:2017jka}. The relevant ratio is traditionally denoted by $R_0$:
\begin{align}
R_0(w) &= \frac{P_1(w)}{A_1(w) },\nn
\end{align}
and of course   $R_0(w)\to 1$ in the heavy quark limit. 
We use the updated NLO HQET calculation for $R_0$ \cite{Bernlochner:2017jka}, include gaussian uncertainties from the
QCD sum rules parameters and $m_b$ in the same way as \cite{Bernlochner:2017jka}, and  in addition we assign to the NLO HQET result a 15\%  
uncertainty from higher order corrections. 
This corresponds to over 30\% uncertainty on $R_{\tau,2}(D^*) $.
Here and in the remainder of this section all the errors are meant to be gaussian errors and are combined in quadrature whenever appropriate: this differs from what we did in Sec.III where we were looking for absolute bounds. In addition, we also impose strong
unitarity constraints on the parameters of the $z$-expansion for $P_1$, see Fig.~\ref{fig:envelopes}: this moves the central value slightly off the one computed using the central values in Table V and reduces somewhat the uncertainty.

Using our fit with LCSR and strong unitarity bounds, we find 
\begin{align}
R_{\tau,1}(D^*) 	  &= 0.232\qquad 
R_{\tau,2}(D^*) 	   = 0.026  \,,\nn\\
R(D^*) 		  &=  0.258 (5) (^{+8}_{-7}) 
  \,,\label{1a}
 \end{align}
where the first error refers to the $B\to D^*\ell\nu$ fit parameters and the full parametric 
uncertainty of $R_0$, while 
the second one is related to the 15\% uncertainty due to higher order corrections to $R_0$. 
The contribution of $R_{\tau,2}(D^*)$ to the final result is about 10\%. 
The uncertainty on $P_1$, which affects only $R_{\tau,2}(D^*)$, has therefore a comparably small impact
on the total  uncertainty of the  SM prediction of $R(D^*)$. It turns out, however, 
that this is the largest single source of uncertainty.
The results obtained with the  fit without LCSR and with strong unitarity bounds are
very similar,
\begin{align}
R_{\tau,1}(D^*) 	  &= 0.232 \,, \qquad 
R_{\tau,2}(D^*) 	   = 0.025 \,,\nn\\
R(D^*) 		  &= 0.257 (5) (^{+8}_{-7})  
\label{1b}\end{align}
where the two errors have the same meaning as in (\ref{1a}).
Combining  all errors in quadrature we end up with $R(D^*) =0.258(^{+10}_{-9})$ and $0.257(^{+10}_{-8})$, respectively. 
These results  agree well with those obtained in 
Ref.\,\cite{Bernlochner:2017jka} using the same normalization to $A_1$, except for the uncertainty due to higher order corrections 
which is not considered there.  They are also compatible with $R(D^*) =0.252\pm 0.003$ \cite{Fajfer:2012vx}
which has been used so far as reference SM prediction in most papers on the subject.

\subsubsection{Normalizing $P_1$ to $V_1$}\label{RDV1}
Let us now proceed to compute $R(D^*)$ in the second way. Only the calculation of 
$R_{\tau,2}(D^*)$ is different from the above derivation.
Here we use the precise determination of $V_1(w) $ from experimental $B\to D\ell\nu$
data and lattice QCD calculations of \cite{Bigi:2016mdz}.
In particular, with the BGL $N=2$ parametrization of  $V_1$ 
and our fit with LCSR and strong unitarity bounds we get 
\begin{align}
R_{\tau,1}(D^*)   &= 0.232\,, \qquad 
R_{\tau,2}(D^*)    = 0.036\,, \nn\\
R(D^*) 		  &= 0.268 (^{+9}_{-8}) (^{+12}_{-10})     \,,\label{2a}
 \end{align}
where the  first error comes from parametric and fit uncertainties, and the second
one from the 15\% higher orders error. Using instead our fit 
  without LCSR input one gets 
\begin{align}
R_{\tau,1}(D^*)   &= 0.232\,, \qquad 
R_{\tau,2}(D^*)    = 0.038\,,  \nn  \\
R(D^*) 		  &= 0.270 (^{+9}_{-8}) (^{+12}_{-10})  \,. \label{2b}
 \end{align}
The values of $R(D^*) $ in Eqs.(\ref{2a},\ref{2b})   are substantially higher and have a larger uncertainty than those obtained with the first method, although they are compatible  within errors. The higher value of $R(D^*) $ is mostly due to the large difference, already noticed in Sec.~II, between the NLO HQET and the lattice QCD predictions for $A_1(1)/V_1(1)$, see Eqs.\,(\ref{lat},\ref{ratiohqe}). 
A lattice QCD determination of the form factor $P_1$, even only at zero-recoil,
would drastically decrease the uncertainty in $R(D^*) $.
 
 \subsubsection{Enforcing a constraint at $q^2=0$ }
 The fits presented in sec.~III allow for a 5\% determination of $A_5$ at the endpoint $w=w_{max}$. This is outside the physical range for the semileptonic decay to taus, see
 (\ref{wtaumax}), but the relation  $P_1(w_{max}) = A_5(w_{max})$ still constrains $P_1(w)$ significantly. We will now use only the fit with strong unitarity bounds and LCSR, which gives
 \begin{align}
 A_5(w_{max})&=0.545\pm 0.025 .\label{end5}
 \end{align}
This is significantly lower than $P_1(w_{max})\simeq 0.69$ obtained using the normalization 
to $V_1$ considered in the previous subsection, and also lower than the $P_1(w_{max})\simeq 0.62$ obtained normalizing $P_1$ to $A_1$. 
For what concerns the value at zero recoil, $w=1$, we can again use $P_1(1)=(P_1/V_1)_{HQET} V_1(1)_{lat}$ or $P_1(1)=(P_1/A_1)_{HQET} A_1(1)_{lat}$, where the lattice values
$V_1(1)_{lat}$ and $A_1(1)_{lat}$ are taken from Eq.\,(\ref{lat0}),
leading to $P_1(1)=1.27(21)$ and $P_1(1)=1.12(18)$. Here we have combined in quadrature the parametric uncertainty with a 15\% theoretical uncertainty.
An intermediate choice consists in using the HQET relation between $P_1$ and the Isgur-Wise function, which is 1 at zero recoil. At the NLO we find
\be
P_1(1)= 1.21 \pm 0.06 \pm 0.18\nn
\ee
where the first error is parametric, and the second corresponds to the 15\%  theoretical uncertainty 
considered above. Using Eq.\,(\ref{P1exp}) this amounts to a  determination of $a_0^{P_1}$,
\be
a_0^{P_1}=0.0595\pm 0.0093,\nn
\ee
which can be combined  with (\ref{end5}) to derive
\be
a_1^{P_1}= -0.318\pm 0.170 - 0.056 a_2^{P_1} \nn
\ee
where the last term must satisfy $|a_2^{P_1}|<1$ and $a_1^{P_1}$ is consistent with strong unitarity for almost any $ a_2^{P_1}$.
Using the last two relations, scanning in the relevant range of $a_2^{P_1}$, and combining the errors in quadrature  we get
\be
R_{\tau,2}(D^*) 	  = 0.028  \,,\quad
R(D^*) 		  = 0.260(5) (6)  \,,\nn
 \ee
where the first error refers to the parametric uncertainty in $R_{\tau,1}(D^*) $ and the second one is related to $P_1$ and parametric uncertainty in $R_{\tau,2}(D^*)$ only. The correlation between the two errors is small.
We observe that the uncertainty is slightly smaller than those of the other methods.
   
\vskip 2mm

The three methods we have employed to compute $R(D^*)$ lead to results which are  consistent within uncertainties. The third method has a slightly smaller error and 
benefits from an important constraint at $q^2=0$ which is not taken into account with the first two methods. In particular, 
 Eqs.~(\ref{2a},\ref{2b}) are likely to somewhat overestimate $R(D^*)$. 
We therefore adopt as our final result the one obtained with the third method,
\be
R(D^*) = 0.260 \pm 0.008, \label{RDstar}
\ee
which still differs $2.6\sigma$ from the experimental world average, but it is higher and 
has an uncertainty almost three times larger than existing estimates\footnote{Only 
Ref.\,\cite{Jaiswal:2017rve}, which appeared together with the first version of this paper, has a larger uncertainty, finding $R(D^*)=0.257(5)$.}.

We apply the same methodology also to the prediction of the longitudinal $\tau$ lepton polarization \cite{Tanaka:1994ay, Fajfer:2012vx, Celis:2012dk,Tanaka:2012nw} 
\begin{align}
P_{\tau} &= \frac{\Gamma^+ - \Gamma^-}{\Gamma^+ + \Gamma^-}\,,
\end{align}
where $\Gamma^{\pm}$ are the integrated decay rates  for definite $\tau$ lepton helicity. One has~\cite{Fajfer:2012vx}
\begin{align}
\frac{d\Gamma^-}{dq^2} &= \frac{G_F^2 \vert V_{cb}\vert^2 \vert \vec{p}\vert q^2}{96 \pi^3 m_B^2} \left(1 - \frac{m_{\tau}^2}{q^2}\right)^2\times \nn\\ 
&\left(H_{--}^2 + H_{++}^2 + H_{00}^2\right)\,, \\
 \frac{d\Gamma^+}{dq^2} &= \frac{G_F^2 \vert V_{cb}\vert^2 \vert \vec{p}\vert q^2}{96 \pi^3 m_B^2} \left(1 - \frac{m_{\tau}^2}{q^2}\right)^2\times \nn\\
&\frac{m_{\tau}^2}{2 q^2}
\left(H_{--}^2 + H_{++}^2 + H_{00}^2 + 3 H_{0t}^2\right)\,,
\end{align}
where in our notation 
\begin{align}
H_{0t} &= m_B \frac{\sqrt{r}(1+r) \sqrt{w^2-1}}{\sqrt{1 + r^2 - 2 w r}} P_1\,,
\end{align}
and $H_{00}, H_{\pm\pm}$ are given in Ref.~\cite{Bigi:2017njr}.
Recently, Belle reported the measurement $P_{\tau} = -0.38 \pm 0.51^{+0.21}_{-0.16}$ \cite{Hirose:2016wfn, Hirose:2017dxl}.
Our SM prediction, independently of the use of LCSR in the fits, is
\begin{align}
P_{\tau} &= -0.47 \pm 0.04\,.
\end{align}

\section{Conclusions}
Unitarity bounds are an essential part of the model independent form factor parametrization 
in semileptonic $B$ decays. They can be made stronger using Heavy Quark Symmetry relations between the $B^{(*)}\to D^{(*)}$ form factors, and are solid and reliable constraints, provided one takes into account conservative uncertainties and recent input
from lattice calculations and experiment.

In this paper we have obtained bounds on the $z$-expansion parameters
of the form factors relevant in the calculation of $B\to D^* \ell \nu$ decays. Since we keep only terms up to $z^2$ in the expansion,  and we generally have lattice QCD information on the first coefficient $a_0$,
the bounds are expressed as allowed regions in the $(a_1,a_2)$  planes for each of the form factors, see Fig.~\ref{fig:envelopes}. 
As lattice QCD calculations extend beyond the zero recoil point, they will soon provide a
relatively precise determination of the slopes of some of the form factors. Our bounds will then become rather strict one-dimensional bounds on the curvature, or on the $a_2$ parameters.

In practice, we have revisited the CLN methodology  20 years later, and used experimental and lattice data to estimate the uncertainties in the HQET relations and to reduce the errors.
Unlike CLN, however, we do not provide a simplified parametrization. On the contrary,
our results on unitarity bounds applied to the BGL parametrization should form the basis of a new generation of 
model independent analyses of $B\to D^* \ell \nu$  data at both Belle-II and LHCb.

For what concerns the determination of $|V_{cb}|$,
we confirm and reinforce the conclusions of our recent analysis \cite{Bigi:2017njr}. The present world average of the exclusive determination of 
$|V_{cb}|$ \cite{Amhis:2016xyh}  relies on the CLN parametrization, but does not include
a reliable estimate of the related theoretical uncertainties and is likely to be biased. 
Although the strong unitarity bounds have  important consequences
on the determination of $|V_{cb}|$ and reduce its value  by about 2\%,
our fits to recent Belle's and lattice data (complemented by the
world average for the $B^0\to D^{*+} \ell\nu$ branching ratio) show a large persisting 
difference  (3.5-5\%)  in the value of $|V_{cb}|$ extracted using the BGL and CLN parametrizations. As already observed in \cite{Bigi:2017njr}, it is possible that such a large difference  is accidentally related to the only Belle data we could analyse for the  $B\to D^* \ell \nu$ differential distributions, and that future global averages of 
Babar and Belle data will lead to a smaller difference between the CLN and BGL fits. However, our approach now includes HQET constraints  with realistic uncertainties and 
improves on the CLN parametrization  in several ways.
Our final results for $|V_{cb}|$ 
are consistent with the inclusive determination but the error is significantly larger, about 3\% instead of 1.5\%. 

We have also reconsidered the SM prediction of $R(D^*)$ in the light of the above results.  Our analysis points to a higher central value and a significantly larger theoretical error than found in previous analyses
\cite{Fajfer:2012vx,Bernlochner:2017jka}.
Our final result is reported in Eq.\,(\ref{RDstar}) and its uncertainty is dominated by the uncertainty in the normalization of the  $P_1$ form factor, which will  be certainly reduced
by future lattice QCD calculations.  
Although we find that its significance is slightly reduced, 
this intriguing flavour anomaly remains a challenge for model builders.

\vspace{2mm}

{\bf Acknowledgements.}
 We are grateful to Marcello Rotondo, Soumitra Nandi, Matthias Neubert, and Christoph Schwanda  for useful discussions.
 
\bibliography{draft.bib}

\end{document}